\documentclass[graybox]{svmult}

\usepackage{helvet}
\usepackage{hyperref}

\usepackage[T1]{fontenc} % if needed
\usepackage{bm}% bold math
\usepackage{url}
\DeclareMathAlphabet{\pazocal}{OMS}{zplm}{m}{n}
\usepackage{braket}
\usepackage{amsmath}
\usepackage{amssymb}
\usepackage{leftidx}
\usepackage[colorinlistoftodos]{todonotes}
\usepackage{bm}
\usepackage{slashed}
\usepackage{calrsfs}
\usepackage{cancel}
\usepackage{dirtytalk}
\usepackage{upgreek}
\usepackage{comment}
\usepackage{xcolor}
\usepackage{ulem}
\usepackage{tensor}

\renewcommand{\emph}{\textit}% selects Helvetica as sans-serif font
\usepackage{courier}        % selects Courier as typewriter font
\usepackage{type1cm}        % activate if the above 3 fonts are
\DeclareUnicodeCharacter{2009}{\,}
\usepackage{makeidx}         % allows index generation
\usepackage{graphicx}        % standard LaTeX graphics tool
\usepackage{multicol}        % used for the two-column index
\usepackage[bottom]{footmisc}% places footnotes at page bottom

\begin{document}

\title*{Unitary  SUSY for the chiral graviton and chiral gravitino in de  Sitter spacetime}
% Use \titlerunning{Short Title} for an abbreviated version of
% your contribution title if the original one is too long
\author{Atsushi Higuchi and Vasileios A. Letsios}
% Use \authorrunning{Short Title} for an abbreviated version of
% your contribution title if the original one is too long
\institute{Atsushi Higuchi \at Department of Mathematics, University of York\\Ian Wand Building, Deramore Lane, York, YO10 5GH, United Kingdom, \email{atsushi.higuchi@york.ac.uk}
\and Vasileios A. Letsios (speaker) \at Physique de l’Univers, Champs et Gravitation, Université de Mons – UMONS \\
Place du Parc 20, 7000 Mons, Belgium,\, \email{vasileios.letsios@umons.ac.be}}
\maketitle

\abstract{It is commonly believed that a unitary supersymmetric quantum field theory (QFT) involving graviton and gravitino fields on fixed 4-dimensional de Sitter spacetime ($dS_4$) cannot exist due to known challenges associated with supersymmetry (SUSY) on
spaces with positive cosmological constant. In this talk, we contradict this expectation
by presenting a new unitary supersymmetric QFT on fixed $dS_4$ : the free supersymmetric theory of the chiral graviton and chiral gravitino fields. The theory overcomes
the known obstacles to unitary global SUSY on de Sitter because the commutator between two SUSY transformations closes on the conformal algebra $so(4,2)$ rather than
the de Sitter algebra $so(4,1)$. Crucially, the $so(4,2)$ symmetry is realised through unconventional conformal-like transformations.  Based on arxiv:2503.04515.}

\section{De Sitter space vs Supersymmetry}
\label{sec:1}

The motivation to study physics in de Sitter (dS) spacetime stems from the relevance of dS spacetime to the theory of Inflation \cite{Baumann}, as well as from its relevance to the current spatially accelerating epoch of our Universe \cite{PlanckCollab, Suzuki, SDSS, Cole}.

 The four-dimensional dS spacetime ($dS_{4}$) is the solution of  vacuum Einstein equations with positive cosmological constant $\Lambda$ and maximal symmetry 
 \begin{equation}
    R_{\mu \nu}-\frac{1}{2} g_{\mu \nu} R + \Lambda  g_{\mu \nu}=0.
\end{equation}
The cosmological constant is $\Lambda = 3 \,\mathcal{R}^{-2}$ ($\mathcal{R}$ is the dS radius). We will adopt units with $\mathcal{R}=1$, while we choose the mostly plus metric signature. The isometry group of  $dS_{4}$  is $SO(4,1)$. Interestingly, there is no generator of time translation symmetry in the dS algebra $so(4,1)$. A convenient coordinate system, which covers the whole dS manifold, corresponds to the global coordinates. In these coordinates, the  $dS_{4}$ line element is
\begin{equation}\label{eq:global metric}
   ds^{2}= - dt^{2}+  \cosh^{2}{t}\,d\Omega^{2},~~~~~~(S^{3}~\text{spatial slices}) ,
\end{equation}
where $t \in \mathbb{R}$. A point on $S^3$ will be denoted as $\bm{\theta}$.

Supersymmetry (SUSY) is a powerful theoretical tool, and is well-understood in anti-de Sitter and Minkowski spacetimes. However, formulating theories with unbroken unitary SUSY in dS space is challenging. Let us first discuss the conceptual problem, and then delve into mathematical details.
  In each of the cases of Minkowski and anti-de Sitter spacetimes, there exists a Grassmann-even positive conserved quantity, identified with the generator of time translations. This quantity appears by taking the `square' of Grassmann-odd generators (supercharges), $Q_{A}$, and summing over spinor indices, as\footnote{For the sake of the discussion, we take $A$ to be  a Dirac spinor index.}: 
 $$\sum_{A} \{ {Q}_{A}, {Q}^{A \, \dagger}   \} \propto \text{positive conserved charge}.$$
 However, in the case of dS spacetime,  no such positive conserved  charge  exists in $so(4,1)$. 
  Let us now  give some more details on the two main problems concerning unitary unbroken SUSY in the presence of positive $\Lambda$:
\\
$\bullet$ \textbf{Non-unitarity of rigid SUSY on a dS background.}  This problem manifests itself at the level of representation theory of the dS superalgebra \cite{Lukierski, Nieuwenhuizen}. The dS superalgebra, apart from the ten dS generators of $so(4,1)$, contains spinorial supercharges ${Q}_{A}$. As shown in \cite{Lukierski, Nieuwenhuizen},  from the structure of the algebra it follows that $\sum_{A} \{ {Q}_{A}, {Q}^{A \, \dagger}   \} = 0$. This means that  all non-trivial representations of the dS superalgebra must have indefinite norm. One can arrive at the same representation theoretic result by doubling the number of supercharges and introducing a symplectic Majorana condition for them \cite{Nieuwenhuizen}. Thus, any supersymmetric QFT on a $dS_{4}$ background, enjoying invariance under the dS superalgebra, is non-unitary.\footnote{Similar results hold for the super-extensions of  $so(3,1)$ and $so(5,1)$ \cite{Lukierski}. However, $so(2,1)$ is an exception, as unitary representation of its super-extension exist \cite{Lukierski} (see also \cite{Anninos_SUGRA}).}
  \\
    $\bullet$ \textbf{Non-unitarity of dS Supergravity.} A $N=2$ Supergravity action with  positive $\Lambda$ was given in \cite{Nieuwenhuizen}. The field content of the theory is: a real vierbein, a real graviphoton, and two symplectic Majorana gravitini.  Although the action is real and invariant under local SUSY, the theory is non-unitary because of the wrong relative sign in the kinetic terms of the graviton and graviphoton.\footnote{On the other hand, dS Supergravity  with broken SUSY exists, and it has no unitarity problems \cite{dS_SUGRA_broken}. Also, note that the supersymmetric higher-spin (Vasiliev-type) gravity theory with positive cosmological constant seems  to be able to bypass the non-unitarity obstacles for unbroken SUSY - see \cite{Hertog, Anninos-Letsios, Sezgin}.}
\\
\\
  \textbf{How  can unitarity of rigid SUSY on a $dS$ background be achieved?} Consider conformal theories on a fixed $dS_{4}$ background, which do not only have $so(4,1)$ symmetry but also enhanced conformal symmetry , $so(4,2) \supset$ $so(4,1)$. The supersymmetrisation of such theories enjoys invariance under  the super-conformal algebra, which does not contain the dS superalgebra as its subalgebra. Now the quantity $\sum_{A} \{ {Q}_{A}, \ {Q}^{A\,\dagger}  \} \geq 0$ corresponds to a positive conserved Grassmann-even charge of $so(4,2)$, associated with ``conformal energy'' , and unitarity is indeed possible \cite{Anous, Dobrev}.

 However, there is a drawback in studying (super-)conformal theories on a dS background, which is conformally flat: one might argue that such theories do not carry intrinsic de Sitterian information, since they can be mapped to theories on flat spacetime.
Nevertheless, a valuable representation theoretic lesson has been learnt: a QFT on a dS background that enjoys a symmetry larger than dS symmetry can, in fact, admit unitary rigid supersymmetry.
Motivated by this observation, in \cite{Higuchi-Letsios}—the paper whose main results are presented in this proceedings contribution—we constructed a new unitary supersymmetric QFT on $dS_{4}$, but outside the superconformal framework.
In particular, we extended 
$so(4,1)$  to $so(4,2)$, though not via the usual conformal transformations. In our construction, the $so(4,2)$ algebra arises from 
$so(4,1)$ transformations together with unconventional conformal-like generators \cite{Letsios_conformal-like}.

\subsection{Main result} 
The main  result of our paper \cite{Higuchi-Letsios} is that there is a new unitary supersymmetric QFT on $dS_{4}$, which includes a chiral version of the fields of the Supergravity multiplet: the free \textit{supersymmetric theory of the  chiral graviton and chiral gravitino fields}.
We use the term `chiral' to refer to gauge potentials whose
field strengths are  anti-self-dual, and thus complex (a chiral gauge potential has one complex propagating degree of freedom). The theory does not seem to be associated with a local action, as it is not clear how to impose the anti-self-duality condition  off-shell.
This new supersymmetric theory    avoids the unitarity no-go theorem mentioned above because the commutator of two SUSY variations  closes  $so(4,2) \bigoplus u(1)$ instead of $so(4,1)$. 
As mentioned earlier, the $so(4,2)$ algebra arises from 
$so(4,1)$ transformations together with unconventional conformal-like transformations. 

In the rest of this article, we will discuss: the chiral gravitino field (Section \ref{Sec: gravitino}), the chiral graviton field (Section \ref{Sec: graviton}), and finally, the unitary representation of SUSY carried by these two fields (Section \ref{Sec:unitary susy}).

%%%%%%%%%%%
\section{The free  gravitino gauge potential on $dS_{4}$:   locality of the hermitian action and unitarity lead to chirality}\label{Sec: gravitino}
Consider the free gravitino field on $dS_{4}$. This is described by a vector-spinor, $\Psi_{\mu}$,  satisfying the Rarita-Schwinger  equation \cite{Deser, Letsios_announce, Letsios_announce, Letsios_announce_II}
\begin{equation}\label{RS_eqn imag mass}
   \gamma^{\mu \rho \sigma} \left(  \nabla_{\rho}+\frac{i}{2}\gamma_{\rho} \right) \Psi_{\sigma}=0,
\end{equation}
where we have suppressed spinor indices.
The gravitino field is also known as the (strictly) massless spin-3/2 field or spin-3/2 gauge potential. 
Note that the mass parameter is imaginary\footnote{The free gravitino gauge potential on $AdS_{4}$ has a real mass parameter: $\gamma^{\mu \rho \sigma} \left(  \nabla_{\rho}+\frac{1}{2}\gamma_{\rho} \right) \Psi^{(AdS)}_{\sigma}=0$.}.
However, the field is not tachyonic, it propagates on the light-cone \cite{Anninos-Letsios, Deser}.  
The field equation~(\ref{RS_eqn imag mass}) is invariant under infinitesimal gauge transformations 
\begin{equation}\label{gauge_transf_spin3/2}
    \delta\, \Psi_{\mu}= \left( \nabla_{\mu}+\frac{i}{2} \,\gamma_{\mu} \right) \lambda,
\end{equation}
where $\lambda$ is a spinor gauge function.  The gravitino field strength, which is invariant under such gauge transformations, is
\begin{equation}\label{eq: gravitino field strength}
F_{\rho \sigma}=\nabla_{[\rho}\Psi_{\sigma]}+\frac{i}{2}\gamma_{[\rho}  \Psi_{\sigma]}.
\end{equation}
The field equations (\ref{RS_eqn imag mass}) do \textbf{not} admit Majorana solutions \cite{Higuchi-Letsios}. Here we consider complex (Dirac) solutions.

It is convenient to fix the transverse and $\gamma$-traceless gauge ($TT$ gauge). The field equations and the $TT$ gauge conditions are 
\begin{equation}\label{RS equation TT gauge}
     \slashed{\nabla} \Psi^{(TT)}_{\mu}=-i \Psi^{(TT)}_{\mu},\hspace{4mm} \gamma^{\mu} \Psi^{(TT)}_{\mu} = \nabla^{\mu} \Psi^{(TT)}_{\mu}=0.
\end{equation}
These equations are preserved under  gauge transformations (\ref{gauge_transf_spin3/2}) with restricted gauge functions: $\slashed{\nabla} \lambda = -2i \lambda$.
%%%%%%%%%%%%%%%%%%%%%%%%%%%%%%%%%%%%%%%%%%%%%%%%%%%%%%%%%%%
\subsection{Quantisation and anti-self-duality constraint}

To quantise the theory, it is customary to start from a hermitian action functional. However, the conventional Rarita-Schwinger action~\cite{Freedman}
\begin{equation} \label{conventional gravitino action}
S = -\int d^{4}x \, \sqrt-{g}\, \overline{\Psi}_{\mu}\gamma^{\mu \rho \sigma} \left(  \nabla_{\rho}+\frac{i}{2}\gamma_{\rho} \right) \Psi_{\sigma}
\end{equation}
is not hermitian due to the imaginary mass term. Moreover, the action is not gauge invariant. 
However, the alternative action~\cite{Higuchi-Letsios}
\begin{align}\label{gravitino action}
    S_{\frac{3}{2}}= -\int d^{4}x \, \sqrt-{g}\, \overline{\Psi}_{\mu}\gamma^{5}\gamma^{\mu \rho \sigma} \left(  \nabla_{\rho}+\frac{i}{2}\gamma_{\rho} \right) \Psi_{\sigma}
\end{align}
is both hermitian and  gauge invariant. We can thus use action (\ref{gravitino action}) to quantise the theory on a $dS_{4}$ background. However, the appearance of $\gamma^{5}$ in the action will lead to some curious chiral features in the quantum theory. To be specific, unlike gravitini in Minkowski or AdS space, in dS space both physical helicities, $\pm 3/2$, cannot coexist in the same Hilbert space if one insists on locality of the hermitian action and unitarity of the theory. Let us elaborate on this further.

Let us first isolate the propagating degrees of freedom in the classical field and then quantise.  We fix the gauge completely for the classical field as ${\Psi}^{(TT)}_{t} =  g^{\tilde{\mu} \tilde{\nu}} \gamma_{\tilde{\mu}} {\Psi}^{(TT)}_{\tilde{\nu}}=0$, where the indices $\tilde{\mu}, \tilde{\nu}$ are spatial indices. This is a Coulomb-like gauge, akin to the Coulomb gauge used in the case of the photon gauge potential.  The mode solutions of (\ref{RS equation TT gauge}) in global coordinates (\ref{eq:global metric}) have been written down explicitly in \cite{Letsios_announce, Letsios_announce_II, Higuchi-Letsios}. In particular, eqs. (\ref{RS equation TT gauge}) admit physical mode solutions, as well as pure-gauge mode solutions.  We expand the quantum field in terms of physical modes (these have vanishing time component), introducing creation and annihilation operators, as
   \begin{align}\label{mode expansion gravitino full}
   \Psi^{(TT)}_{\tilde{\mu}}(t,\bm{\theta}) & = \sum_{\ell, m} \Bigg(a^{-}_{\ell  m}\underset{\text{-3/2 particle}}{\psi^{(-\ell,m)}_{\tilde{\mu}}}(t,\bm{\theta})  +a^{+}_{\ell  m}\underset{\text{+3/2 particle}}{\psi^{(+\ell,m)}_{\tilde{\mu}}}(t,\bm{\theta}) \nonumber  \\
    &~~~~~~~~~~~~~+b^{-~\dagger}_{\ell m}~\underset{\text{-3/2 anti-particle}}{\psi^{(-\ell,m)C}_{\tilde{\mu}}}(t,\bm{\theta})  +b^{+~\dagger}_{\ell m}~\underset{\text{+3/2 anti-particle}}{\psi^{(+\ell,m)C}_{\tilde{\mu}}}(t,\bm{\theta}) \Bigg) ,
    \end{align}
where $\ell = 1 ,2, ...$ is the angular momentum quantum number on $S^{3}$. The label $m$ denotes collectively other quantum numbers. The modes $\{\psi^{(\pm \ell,m)}_{\tilde{\mu}}(t,\bm{\theta})\}$ are modes of helicity $\pm 3/2$ with positive frequency behaviour in the limit $\ell \gg 1$.  The modes $\{\psi^{(\pm \ell,m)C}_{\tilde{\mu}}(t,\bm{\theta})\}$ are the charge conjugates of $\{\psi^{(\pm \ell,m)}_{\tilde{\mu}}(t,\bm{\theta})\}$.
%%%%%%%%%%%%%%%%%%%%%%%%%%%%%%%%%%%%%%%%%%%%%%%%%%%%%%%%%%%%%%%%%%%%%%%%%%%%%%%%%%

As in the case of Minkowski spacetime, the two sets of fixed-helicity modes, $\{\psi^{(+ \ell,m)}_{\tilde{\mu}}(t,\bm{\theta})\}$ and 
$\{\psi^{(- \ell,m)}_{\tilde{\mu}}(t,\bm{\theta})\}$, separately form irreducible representations of the dS algebra. 
(The two sets of charge conjugated fixed-helicity modes, $\{\psi^{(+ \ell,m)C}_{\tilde{\mu}}(t,\bm{\theta})\}$ and 
$\{\psi^{(- \ell,m)C}_{\tilde{\mu}}(t,\bm{\theta})\}$, form the same direct sum of representations.)
More explicitly,  if $\xi^{\mu}$ is any dS Killing vector, then the corresponding infinitesimal dS transformation is encoded by the  Lie-Lorentz derivative $\mathcal{L}_{\xi}$ \cite{Ortin}. A lengthy calculation  shows that modes of fixed helicity do not mix with each other under dS transformations \cite{Letsios_announce_II, Higuchi-Letsios} 
\begin{align} \mathcal{L}_{\xi}\psi^{({\sigma}\ell,m)}_{\mu}=\sum_{\ell',m'}c^{(\ell,m)}_{\ell',m'}~ \psi^{({\sigma}\ell',m')}_{\mu}\,+\, \text{(TT pure-gauge mode)},~~~{\sigma=\pm}.
\end{align}
However, irreducibility is not enough. To have a consistent quantum theory, we need the representations formed by the modes to be unitary. Let us recall the conditions that have to be satisfied for  a representation to be unitary: existence of a \textbf{positive-definite} inner product with respect to which the generators (Lie-Lorentz derivatives) are \textbf{anti-hermitian}, i.e. the inner product must be dS invariant.

 The dS invariant (and gauge invariant) inner product for any two $TT$ solutions $\psi_{\mu}$ and $\psi'_{\mu}$ is
\begin{align}
   & \left( \psi | \psi' \right)_{axial} = \int_{S^{3}}\sqrt{-g} ~d^{3}\bm{\theta}  ~ \left(\psi_{\tilde{\mu}}(t,\bm{\theta})\right)^{\dagger}~{\gamma^{5}}\psi^{'\tilde{\mu}}(t,\bm{\theta}) .
\end{align}
This inner product is determined from the action (\ref{gravitino action}). We observe the appearance of $\gamma^{5}$ in the inner product - if $\gamma^{5}$ is removed, then it will no longer be time-independent or dS invariant.\footnote{The $\gamma^{5}$ in the dS invariant inner product is needed because the mass parameter is imaginary. In the case of a real mass parameter, the inner product without $\gamma^{5}$ should be used.}
Calculating the inner product for the physical modes, we find \cite{Letsios_announce_II, Higuchi-Letsios}
\begin{align}\label{axial inner product modes}
   & \left( \psi^{(\sigma \ell,m)} | \psi^{(\sigma' \ell',m')} \right)_{axial}  \nonumber \\
   &= \int_{S^{3}}\sqrt{-g} ~d^{3}\bm{\theta}  ~ \left(\psi^{(\sigma \ell,m)}_{\tilde{\mu}}(t,\bm{\theta})\right)^{\dagger}~{\gamma^{5}}\psi^{(\sigma' \ell',m')\tilde{\mu}}(t,\bm{\theta})  \nonumber \\
   &=-\sigma\,\delta_{\sigma \sigma'}\delta_{\ell \ell'}\delta_{mm'},~~\text{where}~\sigma=\pm.
\end{align}
  We similarly find $ \left( \psi^{(\sigma \ell,m)C} | \psi^{(\sigma' \ell',m')C} \right)_{axial} =+\sigma\,\delta_{\sigma \sigma'} \delta_{\ell \ell'} \delta_{mm'}   $. We observe something unexpected: although modes of either helicity should correspond to physical propagating degrees of freedom, the norm is not positive for both helicities. Group theoretically,  we are allowed to use a different inner product for each of the two sets, $\{\psi^{(- \ell,m)}_{\tilde{\mu}}(t,\bm{\theta})\}$ and 
$\{\psi^{(+ \ell,m)}_{\tilde{\mu}}(t,\bm{\theta})\}$. In particular, we can use (\ref{axial inner product modes}) for the former and the negative of (\ref{axial inner product modes}) for the latter. Indeed, by making these choices of different positive-definite inner products, one can show that the modes form a direct sum of discrete series Unitary Irreducible Representations (UIRs) of $so(4,1)$ \cite{Letsios_announce, Letsios_announce_II, Higuchi-Letsios}.   However, the locality of the field theory demands that we use the same inner product for both sets of modes \cite{Higuchi-Letsios}. This creates a unitarity problem: although abstract representation theory allows the existence of the gravitino UIRs of either helicity, the corresponding QFT associated with the action (\ref{gravitino action}) is non-unitary.
  In fact, the indefiniteness of the norm also manifests itself in the anti-commutators: $\{a^{\sigma}_{\ell m}, a^{\sigma'\, \dagger}_{\ell' m'}\}=-\sigma \, \delta_{\ell \ell'}  \delta_{mm'}= - \{b^{\sigma}_{\ell m}, b^{\sigma'\, \dagger}_{\ell' m'}\} $.\footnote{As an alternative approach, one could abandon the requirement for the theory to possess a hermitian and local action. Such a quantisation was carried out for the gravitino in the planar patch of $dS_{4}$ in \cite{Anninos-Letsios}. In this approach, only positive-norm states appear, but the inner product is highly non-local. }
%%%%%%%%%%%%%
\\
\\
\noindent \textbf{Chiral quantisation.} How can we achieve unitarity in the QFT Fock space of the gravitino field? An interesting way out is to impose the (on-shell) \textbf{anti-self-duality constraint} on the field strength
\begin{equation}\label{ASD constrtaint gravitino}
   \widetilde{F}_{\mu \nu}\equiv \frac{1}{2}\varepsilon_{\mu \nu \rho \sigma} F^{\rho \sigma} = -i~ F_{\mu\nu} .
\end{equation}
We denote the field strength that satisfies this constraint as ${F}^{-}_{\mu \nu}$.
Interestingly, the anti-self-duality constraint {removes all negative norm states} (i.e. half of the helicities) from the QFT Fock space. This constraint is also consistent with dS invariance and microcausality of the theory - see \cite{Higuchi-Letsios} for details.
 The completely gauge-fixed {quantum chiral gravitino potential}, the field strength (\ref{eq: gravitino field strength}) of which satisfies the anti-self-duality constraint (\ref{ASD constrtaint gravitino}), has the following mode expansion:
   \begin{equation} \label{mode expansion gravitino chiral}
       \Psi^{(TT){\bm{-}}}_{\tilde{\mu}}(t,\bm{\theta})  =    
 \sum_{\ell, m} \Bigg(a^{-}_{\ell \,m}{\psi^{({\bm{-}}\ell ,m)}_{\tilde{\mu}}}(t,\bm{\theta})+b^{+~\dagger}_{\ell\,m}~{\psi^{({\bm{+}}\ell,m)C}_{\tilde{\mu}}}(t,\bm{\theta}) \Bigg) .
   \end{equation}
  The vacuum $ |0\rangle_{\frac{3}{2}}$ is the state annihilated by all annihilation operators.
The two sets of modes $\{\psi^{(- \ell,m)}_{\tilde{\mu}}(t,\bm{\theta})\}$ and 
$\{\psi^{(+ \ell,m)C}_{\tilde{\mu}}(t,\bm{\theta})\}$ in the expansion (\ref{mode expansion gravitino chiral}) form a direct sum of discrete series UIRs of $so(4,1)$ with the same positive-definite inner product (\ref{axial inner product modes}).
However, it is not clear how to achieve the helicity splitting  appearing in (\ref{mode expansion gravitino chiral}) at the level of the action.  Nevertheless, this QFT is consistent with $so(4,1)$ unitarity at the level of the field equations.

%%%%%%%%%%%%%%%%%%%%%%%%%%%%%%%%%%%%%%%%%%%%%%%%%%%%%%%%%%%%%%%%%%%%%%%%%%
\subsection{Conformal-like symmetry for the gravitino}

   The gravitino gauge potential does not only enjoy $so(4,1)$ symmetry. It also enjoys a larger $so(4,2)$ conformal-like symmetry \cite{Letsios_conformal-like, Higuchi-Letsios}. This symmetry appears for both the chiral and the non-chiral fields.
   The new {conformal-like transformations} are generated by the five conformal Killing vectors, $V^{\mu}$, of $dS_{4}$ that are \textbf{not} themselves Killing vectors. We call these `genuine conformal Killing vectors'. The genuine conformal Killing vectors are expressed as gradients of scalars $V_{\mu} = \partial_{\mu}f_{V}$, and they satisfy `half' of the conformal Killing equation: $\nabla_{\mu}V_{\nu} = g_{\mu \nu} \nabla^{\rho} V_{\rho}~/~4.$ \cite{Letsios_conformal-like, Higuchi-Letsios}, as well as the full conformal Killing equation.

The conformal-like transformations, generated by genuine conformal Killing vector $V^{\mu}$, have the following form \cite{Letsios_conformal-like}:
\begin{align}\label{eq" conf-like trans gravitino}
    \mathcal{T}_{V}&\Psi^{(TT)}_{\mu} 
    =  ~\gamma^{5}\Big( V^{\rho}\nabla_{\rho}\Psi^{(TT)}_{\mu}+ i\,  V^{\rho}\gamma_{\rho} \Psi^{(TT)}_{\mu}  -i\, V^{\rho}\gamma_{\mu} \Psi^{(TT)}_{\rho}  \nonumber\\ 
   & + \frac{3}{2} \frac{\nabla_{\rho}{V}^{\rho}}{4} \, \Psi^{(TT)}_{\mu}\Bigg)-\frac{2}{3} \left( \nabla_{\mu} +\frac{i}{2}\gamma_{\mu}\right) \gamma^{5}\Psi^{(TT)}_{\rho}V^{\rho}.
\end{align}
The transformations $\mathcal{T}_{V}$ map solutions of the field equation to other solutions. They are also off-shell symmetries of the action (\ref{gravitino action}). Moreover, they are symmetries for the full non-gauge-fixed theory.
 The transformations (\ref{eq" conf-like trans gravitino})  are \textbf{not}  conventional infinitesimal conformal transformations, and their geometric interpretation is unclear.  The transformation induced on the field strength under (\ref{eq" conf-like trans gravitino}) is a product of a duality rotation times a conventional infinitesimal conformal transformation (times a factor of $i$) \cite{Letsios_conformal-like}.

The five conformal-like generators (\ref{eq" conf-like trans gravitino}), together with the ten infinitesimal dS transformations $\mathcal{L}_{\xi}$, form the algebra $so(4,2)$ (up to gauge transformations) \cite{Letsios_conformal-like, Higuchi-Letsios}.
The two sets of modes that appear in mode expansion of the chiral gravitino (\ref{mode expansion gravitino chiral}), $\{\psi^{(- \ell,m)}_{\tilde{\mu}}(t,\bm{\theta})\}$ and 
$\{\psi^{(+ \ell,m)C}_{\tilde{\mu}}(t,\bm{\theta})\}$, separately form UIRs of $so(4,2)$. Thus, the QFT Fock space of the chiral gravitino field (\ref{mode expansion gravitino chiral}) furnishes UIRs of both $so(4,1)$ and $so(4,2)$.

%%%%%%%%%%%%%%5begin{frame}{Gravitino summary}

%%%%%%%%%%%%%%%%%%%%%%%%%%%%%%%%%%%%%%%%%%%%%%%%%%%%%%%%%%%%%%%%%%%%%%%%%%%%%%%%%

%%%%%%%%%%%%%%%%%%%%%%%%%%%%%%%%%%%%%%%%%%%%%%%%%%%%%%
\section{The free (chiral) graviton gauge potential on $dS_{4}$}\label{Sec: graviton}

The unitarity of the graviton field in dS spacetime has been studied extensively in many works, such as \cite{Higuchi_Linearised, Higuchi_forbidden, Higuchi_Instab, STSHS}. 
The action of the dS graviton is found by linearising the Einstein-Hilbert action around a dS background.  The graviton gauge potential is a real massless spin-2 field, $h_{\mu \nu} = h_{(\mu \nu)}$. The linearised Einstein-Hilbert action (after some integrations by parts) is \cite{Higuchi_forbidden}
\begin{align}\label{real graviton action}
    S_{EH}= -\frac{1}{4}\int d^{4}x \, \sqrt{-g}\,  h^{\mu \nu}\, H_{\mu \nu}(h).
\end{align}
We have defined
\begin{align}\label{lin Einstein operator real}
     H_{\mu \nu}(h)\equiv &\, \nabla_{\mu} \nabla_{\alpha}h^{\alpha}_{\nu}+ \nabla_{\nu} \nabla_{\alpha}h^{\alpha}_{\mu}- \Box h_{\mu \nu}+g_{\mu \nu}\,\Box h^{\alpha}_{~\alpha} -\nabla_{\mu} \nabla_{\nu}h^{\alpha}_{~\alpha} \nonumber \\
    &- g_{\mu \nu}\, \nabla^{\alpha}\nabla^{\beta}h_{\alpha \beta} +2\, h_{\mu \nu}+g_{\mu \nu}h^{\alpha}_{~\alpha},
\end{align}
where $h$ does not represent the trace of $h_{\mu \nu}$. The Laplace-Beltrami operator is denoted as  $\Box = g^{\mu  \nu}\nabla_{\mu} \nabla_{\nu}$. 
The field equations are $H_{\mu \nu}(h)=0$.
The action (\ref{real graviton action}) and the field equations are invariant under gauge transformations (linearised diffeomorphisms)
\begin{align}\label{gauge_transf_spin2_real}
    \delta h_{\mu \nu}= \nabla_{(\mu} Z_{\nu)},
\end{align}
where $Z_{\nu}$ is a vector gauge function. 
The field strength, which is invariant under such gauge transformations, corresponds to the linearised Weyl tensor
\begin{align} \label{def:graviton_field-strength}
    U_{\alpha \beta \mu \nu} =\Big( -\nabla_{\mu}\nabla_{[ \alpha}{h}_{\beta]\nu}-g_{\mu [\alpha} {h}_{\beta] \nu}  \Big)-(\mu \leftrightarrow \nu).
\end{align}

The field equations take a simpler form by imposing the transverse-traceless ($TT$) gauge conditions \cite{Higuchi_Linearised}
\begin{align}\label{EOM graviton TT}
  &  \Box h^{({TT})}_{\mu \nu } = 2 h^{({TT})}_{\mu \nu}, \nonumber \\
  & \nabla^{\mu}h^{({TT})}_{\mu \nu} =0,~~ ~h^{({TT})\alpha}_{~ \alpha} = 0.
\end{align}
These equations still enjoy invariance under  gauge transformations (\ref{gauge_transf_spin2_real}) with restricted gauge function: $\Box Z_{\nu} = - 3 Z_{\nu}$, $\nabla^{\nu}Z_{\nu} = 0$. The physical (Bunch-Davies) modes, as well as the pure-gauge modes, of eqs. (\ref{EOM graviton TT}) have been constructed in \cite{Higuchi_Linearised, STSHS}.
%

%%%%%%%%%%%%%%%%%%%%%%%%%%%%%%%%%%%%%%%%%%%%%%%%%%%%%%%%

\subsection{Standard and chiral quantisations}

\textbf{The real graviton.} The quantisation of the real graviton field in the Bunch-Davies vacuum has been carried out in, e.g., \cite{Higuchi_Linearised, Higuchi_Instab}.
To isolate the  {physical} degrees of freedom, we fix the gauge completely by imposing the conditions: ${{h}}^{{(TT)}}_{t \mu} =0$, and $g^{\tilde{\alpha} \tilde{\mu}} \nabla_{\tilde{\alpha}} {{h}}^{{(TT)}}_{\tilde{\mu}  \tilde{\nu}}=0$. In this gauge, the real quantum  graviton gauge potential is expanded in terms of physical Bunch-Davies modes as  \cite{Higuchi_Linearised, Higuchi_Instab}
\begin{align} \label{mode expansion graviton FULL}
{h}^{{(TT)}}_{\tilde{\mu}   \tilde{\nu}}(t, \bm{\theta})  =&  \sum_{L =2}^{\infty}   \sum_{M}  \left( {c}^{-}_{LM}\, {\varphi}^{(- L, \,M)}_{\tilde{\mu} \tilde{\nu}}(t, \bm{\theta})  + {d}^{+ \dagger}_{LM}\,{\varphi}^{(+L, \,M)\star}_{\tilde{\mu} \tilde{\nu}} (t, \bm{\theta}) \right) \nonumber \\
& +  \sum_{L =2}^{\infty}   \sum_{M}  \left( {d}^{+}_{LM}\, {\varphi}^{(+ L, \,M)}_{\tilde{\mu} \tilde{\nu}}(t, \bm{\theta})  + {c}^{- \dagger}_{LM}\,{\varphi}^{(- L, \,M)\star}_{\tilde{\mu} \tilde{\nu}} (t, \bm{\theta}) \right) ,
\end{align}
where $L=2,3,...$ is the angular momentum quantum number on $S^{3}$, while $M$ denotes other quantum numbers.
The mode solutions in the expansion (\ref{mode expansion graviton FULL}) are physical modes and they satisfy  ${\varphi}^{(\pm L, \,M)}_{t \mu}(t, \bm{\theta})=0$, while they also have vanishing spatial divergence and spatial trace. 
The two sets of fixed-helicity modes $\{  {\varphi}^{(- L, \,M)}_{\tilde{\mu} \tilde{\nu}}(t, \bm{\theta})  \}$ and  $\{  {\varphi}^{(+ L, \,M)}_{\tilde{\mu} \tilde{\nu}}(t, \bm{\theta})  \}$, separately form discrete series UIRs of $so(4,1)$ \cite{Higuchi_Linearised}.{
The same direct sum of UIRs is formed by their complex conjugates, $\{  {\varphi}^{(- L, \,M)\star}_{\tilde{\mu} \tilde{\nu}}(t, \bm{\theta})  \}$ and  $\{  {\varphi}^{(+ L, \,M)\star}_{\tilde{\mu} \tilde{\nu}}(t, \bm{\theta})  \}$.}
In the high-frequency limit, $L \gg 1$, the modes $\{  {\varphi}^{(\pm L, \,M)}_{\tilde{\mu} \tilde{\nu}}(t, \bm{\theta})  \}$ behave as positive frequency modes, while their complex conjugates behave as negative frequency modes. The non-vanishing commutators between creation and annihilation operators are
\begin{align}
[  {c}^{-}_{LM} , {c}^{-\dagger}_{L' M' }  ] = \delta_{L   L'}  \delta_{MM'} ,~~~~~~[  {d}^{+}_{LM} , {d}^{+\dagger}_{L' M' }  ] = \delta_{L   L'}  \delta_{MM'}  .
\end{align}
The Bunch-Davies vacuum is defined as the state annihilated by all annihilation operators.

The dS invariant (and gauge invariant) inner product used for the quantisation is the Klein-Gordon inner product. For any two $TT$ solutions, this inner product is defined as
\begin{align}\label{def: KG inner product spin-2}
     \left({ \varphi^{(1)}| \varphi^{(2)}} \right)_{KG} = \frac{i}{{4}} \int_{S^{3}} d\bm{\theta_{3}} \sqrt{-g} \left(\varphi^{(1) \mu \nu *} ~\frac{\partial}{\partial t}\varphi^{(2)}_{\mu \nu} - \varphi^{(2)}_{\mu \nu}~\frac{\partial}{\partial t}\varphi^{(1) \mu \nu *}  \right).
\end{align}
The Klein-Gordon inner product calculated explicitly for the physical modes gives \cite{Higuchi_Linearised}
\begin{align}
    \left({{\varphi}^{(\,\sigma L, \,M)}|{\varphi}^{(\,\sigma' L', \,M')}}\right)_{KG} =  \delta_{\sigma \sigma'}\delta_{L L'}   \delta_{MM'} ,
\end{align}
\begin{align}
  &  \left({{\varphi}^{(\,\sigma L, \,M)\star}|{\varphi}^{(\,\sigma' L', \,M')\star}}\right)_{KG} = - \delta_{\sigma \sigma'}\delta_{L L'}   \delta_{MM'} , \nonumber \\
   &  \left({{\varphi}^{(\,\sigma L, \,M)\star}|{\varphi}^{(\,\sigma' L', \,M')}} \right)_{KG} =0,
\end{align}
where $\sigma, \sigma' \in \{ +, - \}$.
%%%%%%%%5
\\
\\
\noindent \textbf{The chiral graviton.}   To formulate  a supersymmetric theory, we need the same number of fermionic and bosonic degrees of freedom. Since our gravitino is chiral, we are led to {consider the chiral graviton field} $\frak{h}^{-}_{\mu \nu}$. 
 This satisfies the same field equations with $h_{\mu\nu}$, but it is \textbf{complex}. The completely gauge-fixed real graviton (\ref{mode expansion graviton FULL}) and the completely gauge-fixed chiral graviton are related to each other as
 \begin{align}
  {h}^{(TT)}_{\mu \nu}   = \frak{h}^{(TT)-}_{\mu \nu}+ \left(   \frak{h}^{(TT)-}_{\mu \nu}\right)^{\dagger}.
 \end{align}
 The characteristic feature of the chiral graviton is that its field strength (\ref{def:graviton_field-strength}) satisfies the anti-self-duality constraint
 \begin{equation} \label{ASD constrtaint Weyl}
     \widetilde{U}_{\mu \nu \rho \sigma} \equiv  \frac{1}{2} \varepsilon_{\mu \nu}\,^{\alpha \beta}   U_{\alpha\beta\rho \sigma} = -i U_{\mu \nu \rho \sigma}.
 \end{equation}
 We denote the anti-self-dual part of the linearised Weyl tensor as  ${U}^{-}_{\mu \nu \rho \sigma}$.
 The chiral graviton, which  has one complex propagating degree of freedom,  is identified with the gauge potential of the anti-self-dual linearised Weyl tensor. The mode expansion for the completely gauge-fixed chiral graviton is
 \begin{align} \label{mode expansion graviton chiral}
{\mathfrak{h}}^{{(TT)}-}_{\tilde{\mu}   \tilde{\nu}}(t, \bm{\theta})  = \sum_{L =2}^{\infty}   \sum_{M}  \left( {c}^{-}_{LM}{\varphi}^{(- L, \,M)}_{\tilde{\mu} \tilde{\nu}}(t, \bm{\theta})  + {d}^{+\dagger}_{LM}\,{\varphi}^{(+ L, \,M)\star}_{\tilde{\mu} \tilde{\nu}} (t, \bm{\theta}) \right).
\end{align}
The two sets of modes $\{\varphi^{(- L,M)}_{\tilde{\mu} \tilde{\nu}}(t,\bm{\theta})\}$ and 
$\{\varphi^{(+ L,M) \star}_{\tilde{\mu} \tilde{\nu}}(t,\bm{\theta})\}$ in the expansion (\ref{mode expansion graviton chiral}) form a direct sum of discrete series UIRs of $so(4,1)$ with opposite helicity.
As in the case of the chiral gravitino, it is not clear how to achieve the helicity splitting  at the level of the action.
Nevertheless, at the level of field equations, the theory of the chiral graviton is consistent with dS invariance and microcausality - see \cite{Higuchi-Letsios}.

%%%%%%%%%%%%%%%%%%%%%%%%%%%%%%%%%%%%%%%%%%%%%%%%%%%%%%%%%%%%%%%%%%%%%%%%%%%%%%%%%5

%%%%%%

\subsection{Conformal-like symmetry for the graviton}
   As in the case of the gravitino, the graviton gauge potential does not only enjoy dS symmetry. The (real/complex)  graviton also enjoys a larger  conformal-like symmetry \cite{Higuchi-Letsios}. 
   The new {conformal-like transformations} are generated by the five genuine conformal Killing vectors, $V^{\mu}$, of $dS_{4}$.

In the case of the real graviton, the conformal-like transformations have the following form \cite{Letsios_conformal-like}:
\begin{align}\label{eq: conf-like trans REAL GRAVITON}
 {T}_{V}{h}_{\mu\nu} =  \, V^{\rho}\varepsilon_{\rho \sigma \lambda (\mu}   \nabla^{\sigma} {h}^{\lambda }\,_{\nu)} 
\end{align}
and they are symmetries of the real  non-gauge-fixed linearised Einstein equation, but \textbf{not} of the action (\ref{real graviton action}). They are also symmetries of the field equations in the $TT$ gauge. The five conformal-like transformations (\ref{eq: conf-like trans REAL GRAVITON}) together with the ten standard dS transformations form the algebra $so(5,1)$, up to gauge transformations.

In the case of a complex graviton field $\frak{h}_{\mu \nu}$, it is convenient to define the conformal-like transformations by introducing a factor of $i$, as
\begin{align}\label{eq: conf-like trans CHIRAL GRAVITON}
 \mathcal{T}_{V}\frak{h}_{\mu\nu} =i  \, V^{\rho}\varepsilon_{\rho \sigma \lambda (\mu}   \nabla^{\sigma} \frak{h}^{\lambda }\,_{\nu)} .
\end{align}
These are symmetries for both the complex chiral and complex non-chiral fields. Also, the transformations (\ref{eq: conf-like trans CHIRAL GRAVITON}) are symmetries  for both the non-gauge-fixed equations and the equations in the $TT$ gauge. Now, the five conformal-like transformations (\ref{eq: conf-like trans CHIRAL GRAVITON}) together with the ten standard dS transformations form the conformal algebra $so(4,2)$, up to gauge transformations. The two sets of modes that appear in mode expansion  (\ref{mode expansion graviton chiral}), $\{\varphi^{(- L,M)}_{\tilde{\mu}  \tilde{\nu}}(t,\bm{\theta})\}$ and 
$\{\varphi^{(+ L,M)\star}_{\tilde{\mu} \tilde{\nu}}(t,\bm{\theta})\}$, separately form UIRs of $so(4,2)$ \cite{Higuchi-Letsios}. Thus, the QFT Fock space of the chiral graviton  furnishes UIRs of both $so(4,1)$ and $so(4,2)$.

  The conformal-like symmetry will play an important role in the formulation of the supersymmetric theory later. Note that the conformal-like transformations do not have the form of standard infinitesimal conformal transformations. However, the transformation induced on the linearised Weyl tensor under conformal-like transformations  takes a more familiar form. In particular, under (\ref{eq: conf-like trans REAL GRAVITON})  the transformations of the real linearised Weyl tensor is the product of a duality rotation times a conventional infinitesimal conformal transformation.  Under (\ref{eq: conf-like trans CHIRAL GRAVITON}),  the transformation of the complex linearised Weyl tensor is the product of a duality rotation times a conventional infinitesimal conformal transformation times a factor of $i$ \cite{Letsios_conformal-like}.

%%%%%
\section{Unitary SUSY for the chiral graviton—chiral gravitino supermultiplet on $dS_{4}$}\label{Sec:unitary susy}

Now that we have both a bosonic and a fermionic field—the chiral graviton and the chiral gravitino—with the same number of degrees of freedom, we can ask: can one find SUSY transformations that preserve their field equations?
In other words, the question we wish to address is whether the solution space of the chiral fields carries a representation of SUSY. If it does, can the unitarity no-go theorem for rigid SUSY mentioned in the Introduction be circumvented? 

 So far, we know that the pair of chiral fields $(\frak{h}^{(TT)-}_{\mu \nu}, \Psi^{(TT)-}_{\mu})$ carries  $so(4,1)$ and $so(4,2)$ UIRs in the total single-particle Hilbert space 
 \begin{equation*}
     \mathcal{H}=\mathcal{H}_{boson} \bigoplus \mathcal{H}_{fermion}.
 \end{equation*}
 In this total Hilbert space, the {positive-definite and $so(4,2)$-invariant inner products} are $(\cdot|\cdot )_{KG}$ [eq. (\ref{def: KG inner product spin-2})] and $(\cdot|\cdot)_{axial}$ [eq. (\ref{axial inner product modes})], respectively. For later convenience, we note that the inner products are also invariant under $so(4,2) \bigoplus u(1)$, where $u(1)$ refers to infinitesimal phase rotations.

%%%%%%%%%%%%%%%%%%%%%%%%%%%%%%%%%%%%%%%%%%%%%%%

\subsection{Rigid SUSY transformations} Rigid SUSY transformations are generated by Killing spinors $\epsilon$. On $dS_{4}$, the Killing spinor equation is
\begin{equation}
    \nabla_{\mu}\epsilon = -\frac{i}{2}\gamma_{\mu}  \epsilon.
\end{equation}
There are no Majorana solutions to this equation; there are four complex (Dirac) Killing spinor solutions \cite{Anous, Higuchi-Letsios}. As our chiral fields are themselves complex, the lack of Majorana Killing spinors is not an obstacle.
In \cite{Higuchi-Letsios}, we have shown that  the desired rigid SUSY transformations are given by
  \begin{align}\label{rigid SUSY transforms}
   & \delta^{susy}({\epsilon})\Psi^{(TT)-}_{\mu}=\frac{1}{4}  \left( {i}\, \frak{h}^{(TT)-}_{\mu \sigma}\gamma^{\sigma}+\nabla_{\lambda}  \frak{h}^{(TT)-}_{\mu \sigma} \gamma^{\sigma  \lambda}      \right)\epsilon, \nonumber \\
   &\delta^{susy}({\epsilon}) \frak{h}^{(TT)-}_{\mu \nu}={\overline{\epsilon}} {\gamma^{5}}    \gamma_{(\mu}  \Psi^{(TT)-}_{\nu)} + (\text{non-TT gauge transformation}),
\end{align}
where $\epsilon$ are Grassmann-odd Killing spinors. The transformations (\ref{rigid SUSY transforms}) preserve the field equations in the $TT$ gauge.\footnote{The non-TT gauge transformation in the SUSY transformation (\ref{rigid SUSY transforms}) of the graviton has been added to achieve compatibility with the $TT$ gauge - see \cite{Higuchi-Letsios}.} They also preserve the non-gauge-fixed equations for both chiral and non-chiral complex fields \cite{Higuchi-Letsios}. Importantly, the SUSY transformations (\ref{rigid SUSY transforms}) preserve helicity, i.e. they map physical graviton modes of helicity $\pm 2$ to physical gravitino modes of helicity $\pm 3/2$, and vice versa. Also, it has been shown that the SUSY transformations of the  field strengths are consistent with the anti-self-duality constraint \cite{Higuchi-Letsios}.

%%%%%%%%%%%%

\subsection{Structure of superalgebra} 
To determine the structure of the superalgebra, we compute the commutator between two SUSY transformations. We find that the commutator closes on $so(4,2) \bigoplus u(1)$ (plus gauge transformations), as
\begin{align}
 & [\delta^{susy}(\epsilon_{2}), \delta^{susy}(\epsilon_{1}) ]\frak{h}^{(TT)-}_{\mu\nu}\nonumber\\
 &= -\underset{isometry}{\mathcal{L}_{\xi}}\frak{h}^{(TT)-}_{\mu\nu}+ \underset{conf-like}{\mathcal{T}_{V}}\frak{h}^{(TT)-}_{\mu\nu} - \underset{u(1)}{i ~a}~ \frak{h}^{(TT)-}_{\mu\nu} + (\text{pure-gauge term}) ,  
\end{align}
\begin{align}
   & [\delta^{susy}(\epsilon_{2}), \delta^{susy}(\epsilon_{1}) ]{\Psi}^{(TT)-}_{\mu} \nonumber \\
    &= -\underset{isometry}{\mathcal{L}_{\xi}}\Psi^{(TT)-}_{\mu}+ \underset{conf-like}{\mathcal{T}_{V}}{\Psi}^{(TT)-}_{\mu} - \underset{u(1)}{{i} \frac{5}{2}~a}~ {\Psi}^{(TT)-}_{\mu}+ (\text{pure-gauge term})  . 
\end{align}
We observe that conformal-like transformations, (\ref{eq" conf-like trans gravitino}) and (\ref{eq: conf-like trans CHIRAL GRAVITON}), as well as $u(1)$ transformations, have appeared on the right-hand sides. {The transformation parameters are:} ${\xi^{\mu}} = \frac{1}{4}  \bar{\epsilon}_{2} {\gamma^{5}}  \gamma^{\mu} \epsilon_{1} + (c.c)$ (real Killing vector),  ${V^{\mu}}= \frac{1}{4}  \bar{\epsilon}_{2}   \gamma^{\mu} \epsilon_{1}$ $+ (c.c)$ (real genuine conformal Killing vector), and ${a} = \frac{1}{4}   \bar{\epsilon}_{2}  \gamma^{5} \epsilon_{1} +(c.c)$ (real constant scalar, $\nabla_{\mu}a=0$), where $(c.c)$ stands for `complex conjugate'.

%%%%%%%%%%%%%%%%%%%%%%%%%%%%%%%%%%%%%%%%%%%%%%%%%%%%%%%%%%%%%%%%%%%%%%%%%%%%%%%%%%%%%%%%%%%%%%%%%%%%%%%%%%%%%%%%%%%%%%%%%%%%%%%%%%%%%%%%%%%%%%%%%%
%%%%%%%%%%%%%%%%%%%%%%%%%%%%%%%%%%%%%%%%%%%%%%%%%%%%%%%%%%%%%%%%%%%%%%%

\subsection{Unitarity}
 Since the commutator between two SUSY transformations closes on $so(4,2)  \bigoplus$ $u(1)$, the \textbf{unitarity no-go theorem is avoided}.
However, unitarity must still be verified explicitly.
Let us recall the conditions that have to be satisfied for a representation of SUSY to be unitary~\cite{Furutsu}: 
    (1) Positivity of the norm in both the bosonic and  fermionic solution spaces,
(2)  Invariance of the inner products under the generators of the even subalgebra [$so(4,2)$ $\bigoplus u(1)$ in our case],
    (3) SUSY-invariance of the inner products, in the sense that, for any  $TT$ solution   $\psi_{\mu}$  of (\ref{RS equation TT gauge}), and any  $TT$  solution $\varphi_{\mu \nu}$ of (\ref{EOM graviton TT}), the following {equation}  {holds}: 
    \begin{equation}\label{SUSY invariance of scalar product}
       \left({\delta^{susy}(\epsilon)\psi| \psi}\right)_{axial} =  \left({\varphi| \delta^{susy}(\epsilon)\varphi} \right)_{KG}.
    \end{equation}
    Conditions (1) and (2) have already been verified for the physical modes appearing in the mode expansions of the chiral fields [eqs. (\ref{mode expansion gravitino chiral}) and (\ref{mode expansion graviton chiral})].
Condition (3) can likewise be established—after a lengthy calculation; see \cite{Higuchi-Letsios} for details. As a consistency check, we have also constructed the quantum supercharges acting on the Fock space, and we have  shown that the `traced' anti-commutator is positive: $\sum_{A} \{Q_{A}, Q^{A\dagger} \} \geq 0$.
    
    We thus arrive at our main result: the chiral graviton—chiral gravitino  supermultiplet on $dS_{4}$ forms a unitary representation of SUSY.

An important open question is whether a non-linear completion (with local SUSY) of our free theory exists. If such a theory exists,  then its spin-2 sector will not correspond to the true graviton sector of General Relativity, as the 3-graviton coupling cannot be $u(1)$-invariant.

% For figures use
%

\bigskip

\begin{acknowledgement}
It is a pleasure to thank the organisers of the XVI International Workshop on Lie Theory and Its Applications in Physics. V. A. L.  is supported by the ULYSSE Incentive Grant for Mobility in Scientific Research [MISU] F.6003.24, F.R.S.-FNRS, Belgium.  In the early stage of \cite{Higuchi-Letsios}, the work of V. A. L.  was supported by a studentship from the Department of Mathematics at the University of York and a fellowship from the Eleni Gagon Survivor's Trust for research at the Department of Mathematics at King's College London.
\end{acknowledgement}
\end{document}